\documentclass[preprint,12pt, a4paper]{elsarticle}

\journal{SoftwareX}

\usepackage{booktabs}
\usepackage{multirow}
\usepackage{listings, listings-rust}
\usepackage[frozencache=true,cachedir=minted-cache]{minted}
\usepackage{verbatimbox}
\usepackage{xcolor}% or package color
\definecolor{codegreen}{rgb}{0,0.6,0}
\definecolor{codegray}{rgb}{0.5,0.5,0.5}
\definecolor{codepurple}{rgb}{0.58,0,0.82}
\usepackage{ascii}
\usepackage{hyperref}

\lstdefinestyle{mystyle}{
    language=Python,
    commentstyle=\color{codegray},
    keywordstyle=\color{magenta},
    numberstyle=\tiny\color{codegreen},
    stringstyle=\color{codepurple}\scriptsize\asciifamily,
    basicstyle=\scriptsize\asciifamily,
    breaklines=true,                 
    tabsize=2,
    morekeywords={True, False},
    basewidth={.55em}
}

\newcommand{\tsdownsample}{\texttt{tsdownsample}}

\begin{document}

\begin{frontmatter}

\title{\texttt{tsdownsample}: high-performance time series downsampling for scalable visualization}

\author{Jeroen Van Der Donckt} %}
\author{Jonas Van Der Donckt}
\author{Sofie Van Hoecke}

% \cortext[contrib]{Contributed equally}

\address{IDLab, Ghent University - imec, Technologiepark Zwijnaarde 126, 9052 Zwijnaarde, Belgium}

\begin{abstract}
% Effective exploration of large time series is greatly enhanced by interactive line chart visualizations.
Interactive line chart visualizations greatly enhance the effective exploration of large time series.
% Interactive line chart visualizations are essential for effectively exploring large time series datasets. % in many domains, such as healthcare, finance, and manufacturing. 
Although downsampling has emerged as a well-established approach to enable efficient interactive visualization of large datasets, it is 
% To facilitate efficient visualization of such large datasets, downsampling has emerged as a well-established approach.
% However, applying downsampling in conventional Python visualization tools proves to be challenging, as this is generally not an inherent feature and there is no library offering an easy-to-use interface for high-performance implementations of prominent downsampling algorithms.
% However, applying downsampling proves to be challenging, as this is generally 
not an inherent feature in most visualization tools. Furthermore, there is no library offering a convenient interface for high-performance implementations of prominent downsampling algorithms.
To address these shortcomings, we present \tsdownsample{}, an open-source Python package specifically designed for CPU-based, in-memory time series downsampling. Our library focuses on performance and convenient integration, offering optimized implementations of leading downsampling algorithms. We achieve this optimization by leveraging low-level SIMD instructions and multithreading capabilities in Rust. In particular, SIMD instructions were employed to optimize the argmin and argmax operations. This SIMD optimization, along with some algorithmic tricks, proved crucial in enhancing the performance of various downsampling algorithms.
% a Python package that focuses on performance and integrability for CPU-based, in-memory, time series downsampling. Our library provides highly optimized, CPU-centric implementations of leading time series downsampling algorithms by leveriging low-level SIMD instructions and multithreading capabilities of Rust. Especially optimizing the argmin and argmax operations using low-level SIMD instructions in Rust along with some algorithmic tricks, proved crucial to tune the performance of several downsampling algorithms.
%With exceptional performance and scalability, \tsdownsample demonstrates its value as an important tool for facilitating scalable time series visual analysis. 
We evaluate the performance of \tsdownsample{} and demonstrate its interoperability with an established visualization framework. Our performance benchmarks indicate that the algorithmic runtime of \tsdownsample{} approximates the CPU's memory bandwidth. 
This work marks a significant advancement in bringing high-performance time series downsampling to the Python ecosystem, enabling scalable visualization. The open-source code can be found at \url{https://github.com/predict-idlab/tsdownsample}
% We assess the performance of \tsdownsample{} and showcase its integrability with an existing visualization framework, highlighting its real-world significance in facilitating scalable time series visual analysis.
% In summary, this paper presents \tsdownsample, a high-performance Python package for time series downsampling, which provides a scalable solution for efficient visualization of large time series datasets in various domains.
\end{abstract}

\begin{keyword}
%% keywords here, in the form: keyword \sep keyword
time series \sep visualization \sep downsampling \sep Rust \sep Python \sep SIMD

%% PACS codes here, in the form: \PACS code \sep code
%% MSC codes here, in the form: \MSC code \sep code
%% or \MSC[2008] code \sep code (2000 is the default)

\end{keyword} 

\end{frontmatter}

\section{Introduction}\label{sec:introduction}
%% General time series viz intro
Time series are ubiquitous in many domains, such as healthcare, finance, and manufacturing.
This complex data modality can be challenging to comprehend through summary statistics alone, making visualizations a crucial tool for gaining insights, with line charts proving particularly effective for most tasks~\cite{aigner2007visualizing}.
By following the “overview first, zoom and filter, then details on demand” paradigm~\cite{shneiderman2003overviewfirst}, interactive line chart visualizations allow users to quickly and easily understand the data and identify patterns and trends~\cite{walker2015timenotes}. %The human eye has long been considered the ultimate data mining tool~\cite{lin2005visualizing}, making interactive visualizations an indispensable tool for time series analysis.

%% The need for scalability
Many real-world time series datasets are extremely large, encompassing millions or even billions of data points. As a result, there is a pressing need for scalable visualization techniques that are capable of effectively handling such datasets~\cite{bikakis2018big, godfrey2016interactive}. One approach to realize scalable visualization is through utilizing data aggregation techniques such as downsampling, which reduces the number of data points in a time series while preserving its overall shape~\cite{steinarsson2013downsampling, jugel2014m4, aigner2007visualizing}. Downsampling enables faster rendering and more responsive interactions, allowing users to explore large datasets more effectively~\cite{vanderdonckt2022plotly_resampler, liu2014effects}. 
Downsampling algorithms find wide adoption in the time series database domain, with Uber integrating a downsampling function in their M3 metrics platform~\cite{uber_raskin_aggarwal_2018}, and TimeScaleDB offering downsampling as a server-side hyperfunction~\cite{timescaleblog_paganini_2023}.

However, when dealing with very large datasets (billions of data points), the downsampling process itself can become a bottleneck~\cite{van2023minmaxlttb}. This is especially the case in the context of interactive visualizations, which require fast downsampling to minimize latency when interacting with the graph, such as zooming and panning~\cite{liu2014effects}. Moreover, the authors observe that, at the time of writing, there is no Python library offering high-performance implementations of multiple time series downsampling algorithms.
% Many real-world time series datasets are large, containing millions to even billions of datapoints~\cite{TODO}.
% As a result, there is a need for scalable visualization techniques to facilitate the exploration of these datasets.
% One way to achieve scalable visualization is to use data aggregation techniques, such as downsampling~\cite{steinarsson2013downsampling, jugel2014m4}.
% Downsampling is a data reduction technique that reduces the number of datapoints in a time series, while aiming to preserve the overall shape of the data\footnote{Discussing the visual quality of downsampling is out of scope of this paper.}.
% Hence, fewer datapoints have to be visualized, which allows for faster rendering and more responsive interaction.
% However, when dealing with large datasets, the downsampling process itself can become a bottleneck.
% Especially, interactive visualizations require fast downsampling algorithms, as they perform downsampling on interaction events (e.g., zooming, panning).

%  making it important to have scalable downsampling algorithms to ensure rensponsiveness.
% , as it is often performed on the fly, on interaction events (e.g., zooming, panning).
% This is especially the case for interactive toolkits, as downsampling is performed on the fly, on interaction events (e.g., zooming, panning).
% As a result, it is important to have scalable downsampling algorithms to ensure rensponsiveness.

%% Present tsdownsample
Recognizing this challenge, we introduce \tsdownsample{}, an open-source Python toolkit designed for in-memory, CPU-based, time series downsampling, focusing on performance and integrability. 
% As such, to enable effective exploration of large time series datasets, high-performance downsampling algorithms are essential. To address this challenge, we present \tsdownsample, a Python toolkit for in-memory time series downsampling that focuses on performance and integrability. 
\tsdownsample{} provides optimized CPU implementations of the most prominent downsampling algorithms, i.e., EveryNth, MinMax, M4~\cite{jugel2014m4}, LTTB (Largest-Triangle-Three-Buckts)~\cite{steinarsson2013downsampling}, and MinMaxLTTB~\cite{van2023minmaxlttb}.
The algorithms are implemented in Rust, a system programming language known for its performance and memory safety. The Rust code leverages SIMD (Single Instruction Multiple Data) instructions together with some algorithmic tricks and (optionally) multithreading to achieve exceptional performance and scalability. A core component of \tsdownsample{} is our \texttt{ArgMinMax} Rust library, which provides SIMD accelerated argmin and argmax functionality. 
Optimizing these operations for various CPU architectures proved to be crucial, as they form the inner loop of most downsampling algorithms~\cite{van2023datapointselection}. %(given the importance of vertical extrema for visual representativeness of time series downsampling for visualization~\cite{van2023datapointselection}).
\tsdownsample{} is distributed as a \href{https://pypi.org/project/tsdownsample/}{Python toolkit} by publishing the cross-compiled Python bindings for the underlying Rust code for a wide range of operating systems and CPU architectures.
% To facilitate the exploration of large time series datasets, high-performance downsampling algorithms are crucial.
% To that end, we present \tsdownsample, a Python toolkit for time series downsampling, that focuses on performance and integrability.
% \tsdownsample provides highly optimized implementations of downsampling algorithms for time series, including a novel algorithm that is a combination of the Largest Triangle Three Buckets (LTTB) and the Min-Max algorithm.
% \tsdownsample is written in Rust, a systems programming language that is known for its performance and memory-efficiency. 
% The Rust code leverages SIMD instructions, and (optinally) multi-threading to achieve high performance.

%% Evaluation
% The performance of \tsdownsample is evaluated by assessing the memory overhead and the scalability of the downsampling algorithms.
% Our results show that \tsdownsample has no memory overhead and scales linearly with the number of datapoints.
% \tsdownsample even allows for downsampling up to 10 billion datapoints in less than a second, on a consumer-grade laptop (when using multi-threading).
% The integrability of \tsdownsample is threefold; 
% (1) the code is distributed to work on any platform,
% (2) the code is open-source, and 
% (3) the code has a convenient API in Python, enabling asy integration in existing Python-based visualization frameworks.
% We demonstrate the last point by incorporating \tsdownsample in an existing visualization framework, facilitating next-level scalable time series exploration.

In summary, this paper contributes \tsdownsample{}, a high-performance library optimized for CPU that provides downsampling for scalable time series visualization. This library's integrability is demonstrated through its adoption as the downsampling solution in a time series visualization library, which has over 1.5 million installations at the time of writing.

The remainder of this paper is structured as follows. In Section~\ref{sec:software}, we present a description of the software. We then provide an illustrative example in Section~\ref{sec:example}, where we also highlight its integration in an existing toolkit. Finally, we evaluate the performance of \tsdownsample{} in Section~\ref{sec:benchmarks}.

% We evaluate the performance of \tsdownsample{}, measuring the memory overhead and scalability for the implemented downsampling algorithms. In particular, we highlight the advantage of multithreaded execution. 

% The performance of tsdownsample was evaluated by measuring the memory overhead and scalability of the downsampling algorithms. Our results show that tsdownsample has no memory overhead and scales linearly with the number of data points. In fact, when using multithreading, tsdownsample is able to downsample up to 10 billion data points in less than a second on a consumer-grade laptop. In terms of integrability, tsdownsample is designed to work on any platform, is open-source, and has a convenient Python API, making it easy to integrate into existing Python-based visualization frameworks. To demonstrate this, we incorporated tsdownsample into an existing visualization framework, enabling next-level scalable time series exploration.

\section{Software description}\label{sec:software}
\tsdownsample{} is a Python package that utilizes Rust to provide CPU-optimized implementations of downsampling algorithms for time series visualization. To facilitate seamless installation and usage, Python bindings for the underlying Rust code are cross-compiled for various operating systems and CPU architectures, which are distributed as a PyPi package. Installing \tsdownsample{} is simple and can be done via pip by running the following command: \href{https://pypi.org/project/tsdownsample/}{\texttt{pip install tsdownsample}}.

% The Rust code is compiled to a Python extension, which is then distributed to work on any platform.
% The high performance of \tsdownsample is achieved through using SIMD instructions and (optional) multi-threading in Rust. 

% In the following subsections, we will delve further into the performance and user-interface aspect of \tsdownsample{}. 
The following subsections detail the performance and user-interface aspect of \tsdownsample{}.
We will first describe how optimizing argmin and argmax operations in Rust proved crucial for achieving high-performance downsampling. Then, we will discuss the convenient interface that this library provides for all implemented downsampling algorithms.

% Users can install \tsdownsample using pip: \href{https://pypi.org/project/tsdownsample/}{\texttt{pip install tsdownsample}}. %TODO: add pip url

% In the following subsesctions, we will first describe how we optimized the argmin and argmax operations in Rust. We will then explore how this efficient implementation can be utilized in the downsampling algorithms.

\subsection{ArgMinMax}
\label{sec:argmin_argmax}
% The Min-Max, M4, and MinMaxLTTB algorithms all require the argmin and argmax operations.
% These operations are used to find the index of the minimum and maximum values in a given range.
% As these operations are used in the inner loop  of the downsampling algorithms (i.e., for each bin), they are crucial for the performance.
% Hence, we created the ArgMinMax Rust crate, which provides highly optimized implementations of argmin and argmax operations.
% More specifically, ArgMinMax provides SIMD-optimized implementations of argmin and argmax operations (in one function).
% The implementation has the following characteristics:
% \begin{itemize}
%     \item SIMD algorithm is branchless, making the runtime independent of the quality of the branch predictor (which for this task should predict the distribution of the time series data). As a result, the best-case runtime is the same as the worst-case runtime.
%     \item SIMD algorithm implemented for SSE, AVX(2), and AVX512, NEON.
%     \item Runtime cpu feature detection, allowing to select at runtime the best (supported) SIMD implementation for the current CPU.
%     \item Wide support of data types: f16, f32, f64,, i8, i16, i32, i64, u8, u16, u32, u64.
%     \item Memory-efficient: we work on a memory view of the data, which allows us to avoid copying the data.
% \end{itemize}
Given the significance of vertical extrema for ensuring the visual representativeness of time series downsampling~\cite{van2023datapointselection}, it was imperative to optimize the argmin and argmax operations. These operations play a vital role in the inner loop 
% Key to this library was optimizing argmin and argmax operations, which are crucial components to the inner loop 
of the MinMax, M4~\cite{jugel2014m4}, and MinMaxLTTB~\cite{van2023minmaxlttb} algorithms. As such, we developed the ArgMinMax Rust library (also referred to as a crate), which provides a highly efficient and overflow-free implementation of the argmin and argmax operations. These operations return the indices of the minimum and maximum values of an array. Note that the argmin and argmax values are extracted simultaneously within a single pass over the data, as this is mainly a memory-bound task.

The \href{https://crates.io/crates/argminmax}{ArgMinMax crate} includes SIMD-optimized implementations of argmin and argmax for SSE, AVX(2), AVX512, and NEON, and includes runtime CPU feature detection to select the optimal (supported) SIMD implementation for the current CPU. SIMD instructions allow the CPU to perform the same operation on multiple data points simultaneously, providing a significant boost in performance for certain types of operations. The library is SIMD-optimized for a wide range of CPU architectures; x86, x86\_64, arm(v7), and aarch64.
In addition, the ArgMinMax crate supports a wide range of data types (f16, f32, f64, i8, i16, i32, i64, u8, u16, u32, and u64). We further guarantee the library to be memory-efficient, as it operates on a memory view (i.e., a slice) of the data rather than copying it. The SIMD algorithm is also branchless, ensuring that the runtime is independent of the quality of the branch predictor, making the best-case runtime the same as the worst-case runtime.

\subsubsection{argminmax SIMD algorithm}
In code snippet~\ref{listing:simd_algo}, we present the inner loop of the SIMD argmin and argmax algorithm. This algorithm extracts both the argmin and argmax value in a single pass over the data. To do so, we utilize four accumulating SIMD vectors (also referred to as registers). 
Two of these registers maintain the lowest and highest values encountered while iterating over the data in chunks of size \texttt{LANE\_SIZE}. At the end of the iteration, these vectors contain the maximum and minimum values at each position within all seen \texttt{LANE\_SIZE} chunks. 
As a final step, after iterating over all the chunks, the algorithm extracts the minimum and maximum values along with their respective indices from the SIMD vectors (i.e., the horizontal operations).

\begin{listing}[ht]
\begin{lstlisting}[language=Rust]
arr_ptr = arr.as_ptr(); // Array pointer we will increment in the loop
new_index = INITIAL_INDEX; // Index we will increment in the loop

// Initialization of the accumulating SIMD vectors
// Note that for adequate float (NaN) handling this initialization should be different
index_low = INITIAL_INDEX;
values_low = _mm_loadu(arr_ptr);
index_high = INITIAL_INDEX;
values_high = _mm_loadu(arr_ptr);

for _ in 0..arr.len() / LANE_SIZE - 1 { // Iterate over the array in LANE_SIZE chunks
    // Increment the index
    new_index = _mm_add(new_index, INDEX_INCREMENT);
    // Load the next chunk of data
    arr_ptr = arr_ptr.add(LANE_SIZE);
    new_values = _mm_loadu(arr_ptr);

    // Update the lowest values and index
    mask_low = _mm_cmplt(new_values, values_low);
    values_low = _mm_blendv(values_low, new_values, mask_low);
    index_low = _mm_blendv(index_low, new_index, mask_low);

    // Update the highest values and index
    mask_high = _mm_cmpgt(new_values, values_high);
    values_high = _mm_blendv(values_high, new_values, mask_high);
    index_high = _mm_blendv(index_high, new_index, mask_high);
}

// Get the min/max index and corresponding value from the SIMD vectors
(min_index, min_value) = _horiz_min(index_low, values_low);
(max_index, max_value) = _horiz_max(index_high, values_high);
\end{lstlisting}
\caption{The core (inner loop) of the SIMD argminmax algorithm.}
\label{listing:simd_algo}
\end{listing}

The pseudocode in snippet~\ref{listing:simd_algo} closely resembles the Rust code of the ArgMinMax package. It is worth noting that the SIMD instructions, such as \texttt{\_mm\_loadu}, \texttt{\_mm\_cmplt}, \texttt{\_mm\_cmpgt}, and \texttt{\_mm\_blendv}, are generic function names that need to be associated with the corresponding CPU instructions of the various architectures. To achieve this in Rust, we utilized a trait that defines these generic SIMD instructions as functions, similar to C++ templates. For each supported CPU architecture and data type combination in the ArgMinMax package, we have implemented a concrete version of this trait\footnote{Since the lane-size is also influenced by this combination, this trait includes the lane size parameter as well.}.

When the length of the array exceeds the maximum value that the index vector's underlying data type can represent, an (index) overflow will occur. It is this overflow challenge that makes the argmin and argmax operations a much harder problem to SIMD-optimize compared to the min and max operations. As a result, compiling a scalar implementation to vectorized instructions is not trivial or even impossible. As such, it was necessary to manually write the algorithm using SIMD instructions, rather than relying on the compiler for optimized compilation. Notably, the \texttt{polars} library, an exceptionally fast DataFrame library and in-memory query engine that currently exceeds 0.7M monthly installations, adopted our ArgMinMax crate to provide more optimized argmin and argmax operations~\cite{ritchie_vink_2023_7832012}.
For more details on how to implement an overflow-free solution (which requires an additional outer loop), please refer to the open-source code repository available at \href{https://github.com/jvdd/argminmax}{github.com/predict-idlab/argminmax}.

% Note that overflow occurs when the array length exceeds the maximal value for the index vector underlying value data type\footnote{The fact that the accumulating index vector can overflow, makes this a much harder problem to SIMD-optimize (i.e., automatically SIMD compile) than min/max.}. For further details on how to realize an overflow-free implementation (resulting in an additional outer loop) we refer to the \href{https://github.com/jvdd/argminmax}{open-source code base}.

\subsubsection{Optimized implementation for f16 and uints}
In contrast to other data types, most modern CPUs (x86) do not have hardware support for the float16 (f16) data type\footnote{With float16 we refer to IEEE 754-2008 standard binary16, also known as half floating point type~\cite{cornea2009ieee}.}. As a result, programming languages typically support f16 either by upcasting to f32 or by using a software implementation. Both approaches come at the cost of considerable overhead.

\begin{figure}[!tbp]
  \centering
  \begin{minipage}[b]{0.49\textwidth}
    \includegraphics[width=\textwidth]{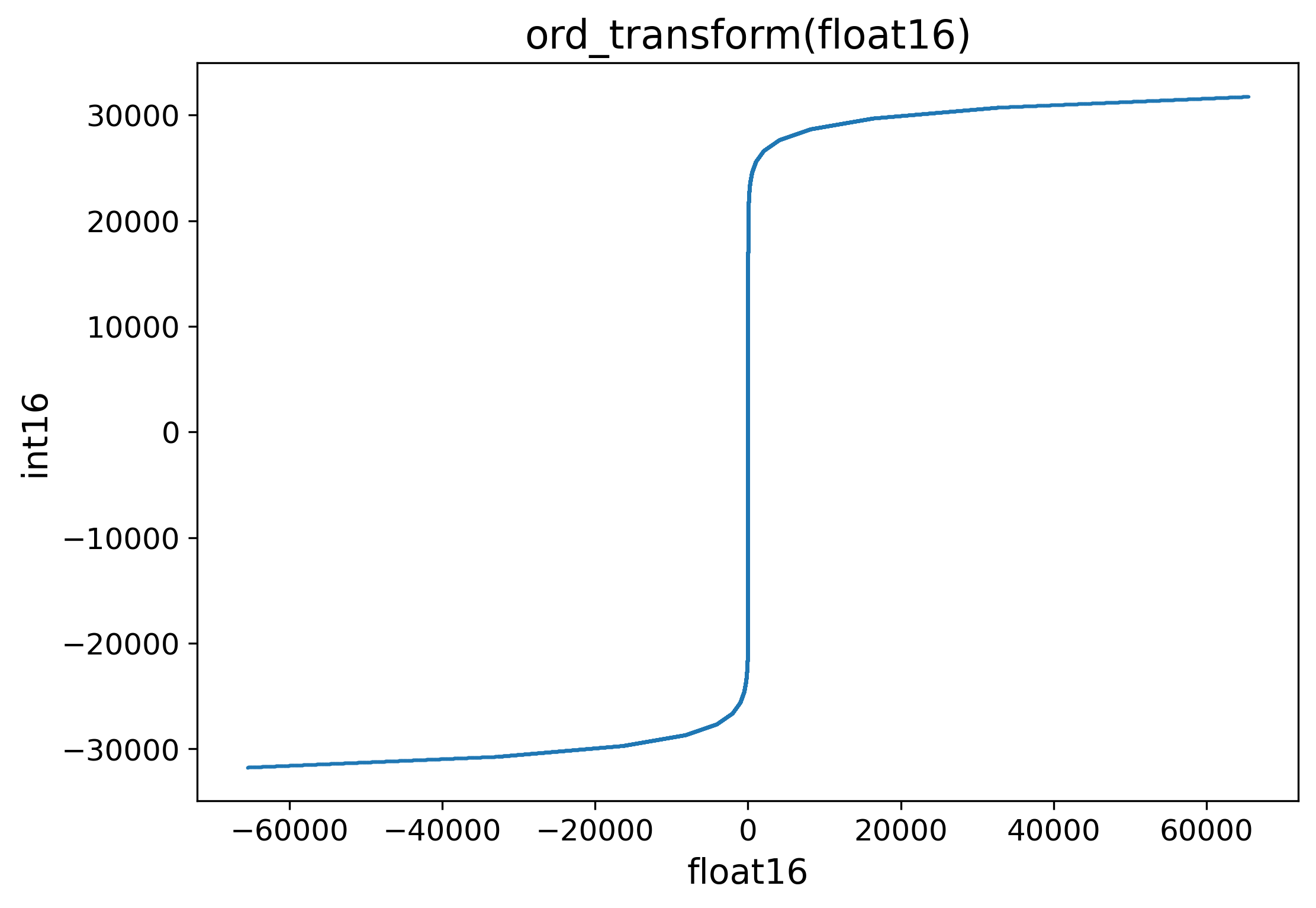}
    \caption{f16 $\rightarrow$ i16}
    \label{fig:f16_to_i16}
  \end{minipage}
  \hfill
  \begin{minipage}[b]{0.49\textwidth}
    \includegraphics[width=\textwidth]{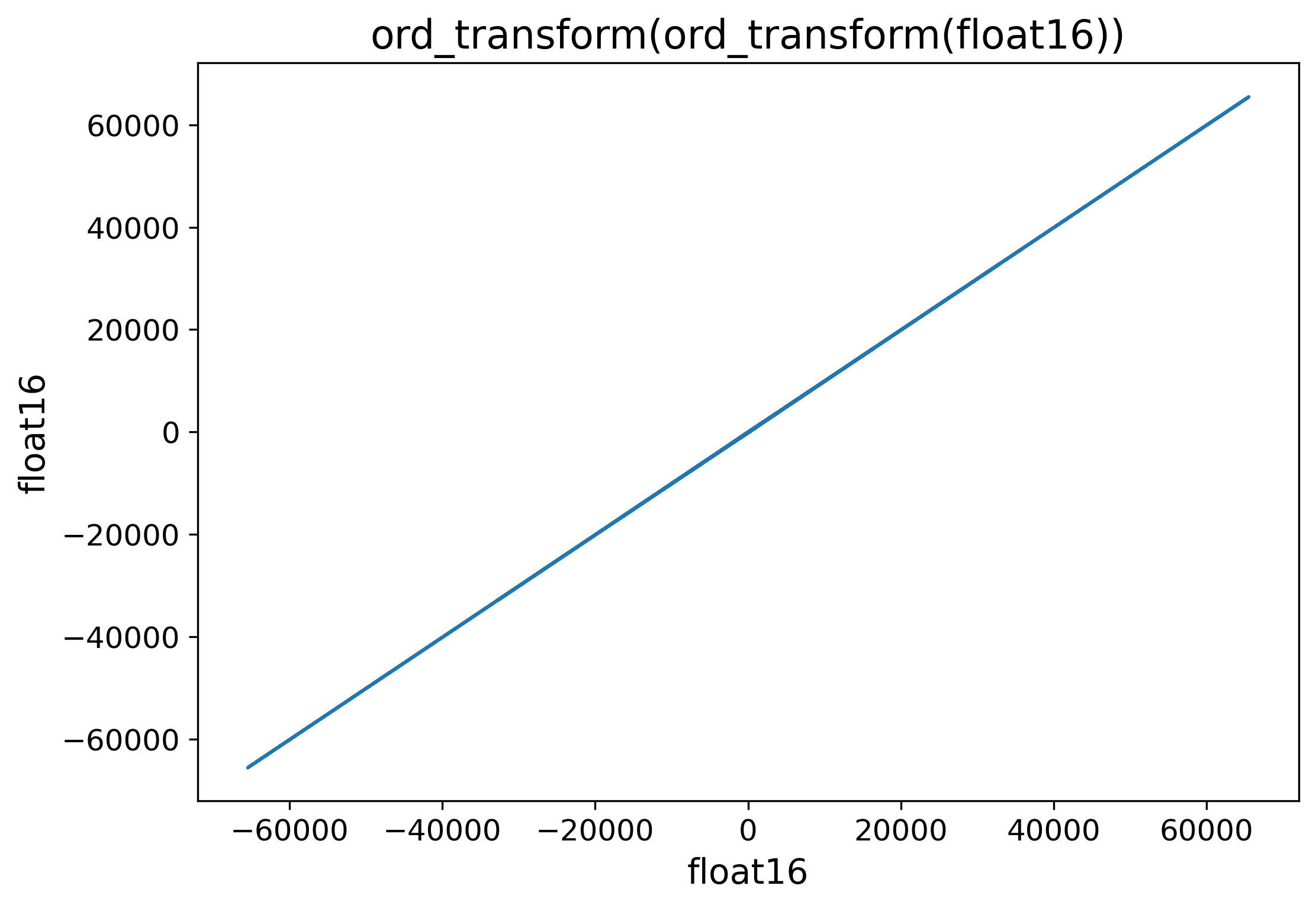}
    \caption{f16 $\rightarrow$ i16 $\rightarrow$ f16}
    \label{fig:f16_to_i16_to_f16}
  \end{minipage}
  % \caption{Value representation oud_transform}
\end{figure}

Instead of applying one of these two approaches, we convert f16 to an ordinal mapping of i16 (which we refer to as i16ord). This allows us to efficiently support f16 data types in the ArgMinMax crate and in the \tsdownsample{} package as a whole.
The mapping preserves the ordinality of the f16 data, as illustrated in Figure~\ref{fig:f16_to_i16}, allowing the use of fast built-in i16 (SIMD) instructions for comparison\footnote{Note that this mapping only makes sense when you are solely interested in comparing values.}. Moreover, the transformation is symmetric, meaning that we can transform the outcome back to f16 (by using the same mapping function) without needing a lookup table, as illustrated in Figure~\ref{fig:f16_to_i16_to_f16}. 
Furthermore, as the transformation only performs binary (bitwise) operations, the overhead is limited, thus making it convenient to implement using SIMD instructions.

The ordinal transformation is performed as follows\footnote{To apply this transformation to a f16 value, we first transmute the f16 value to i16.}:
\begin{center}
    \texttt{ord\_transform(v: i16) = ((v >> 15) \& 0x7FFF) $\oplus$ v)}  
\end{center}

%Then we XOR that value with the  

A final implementation remark is that uint data types (u8, u16, u32, u64) lack SSE, AVX(2), nor AVX512 SIMD instructions. This data type is supported by performing a similar, albeit much simpler, ordinal mapping to integer data type. This mapping leverages the two's complement of signed integers. In particular, the transformation first transmutes the uint as int and then XORs the value with the smallest (i.e., the largest negative) int value. 

\subsection{tsdownsample}
\tsdownsample{} further builds upon the optimizations of the ArgMinMax crate. In particular, the MinMax, M4, and MinMaxLTTB algorithms directly rely on the argmingmax algorithm in their inner loop (i.e., for each bin).  Since these algorithms operate on local heuristics within each bin, they can be easily parallelized in Rust to leverage the processing power of modern multicore CPUs~\cite{heyman2020comparison}. To achieve this, we implemented a multithreaded bin index generator using a search sorted approach, thereby enhancing cache hits through chunked execution. Remark that multithreading is not possible in Python due to the Global Interpreter Lock (GIL) which prevents multiple threads from executing Python bytecodes simultaneously~\cite{beazley2010understanding}. 

Unit testing is conducted to ensure the correctness of the supported downsampling algorithms. Specifically, we verify the consistency of the downsampled (i.e., selected) data points across various data types, downsamplers, and multithreading configurations. Additionally, we compare the Rust implementations to a reference Python implementation, which (although being considerably slower) serves as a benchmark for correctness. Furthermore, we ensure that passing an equally sampled \texttt{x} yields the same output as not specifying the index.
Noteworthy, to capture potential performance regressions when updating the code base, we added performance monitoring to the CI/CD workflow.

% By utilizing SIMD instructions and multithreading, the tsdownsample package is able to achieve exceptional performance and scalability when downsampling large time series datasets.

\subsection{Downsampling interface}\label{sec:downsampling_interface}
\tsdownsample{} aims to provide a convenient interface for the supported downsampling algorithms. Users can interact with these algorithms through Python classes, which act as a thin wrapper around the underlying Rust bindings. These classes abstract the dispatching of data type specific function calls, making the interface easy to use. All classes implement the \texttt{downsample} method, which has the following signature:

\begin{center}
    \texttt{downsample([x], y, n\_out, **kwargs) -> ndarray[uint64]}
\end{center}

This signature first accepts two positional arguments that represent the input values\footnote{This signature design aligns with the convention used in the \texttt{matplotlib.pyplot.plot} method~\cite{hunter2007matplotlib}.}. The first positional argument \texttt{x} is optional and represents the index of the time series values (\texttt{y}). If not provided, it is assumed that the time series values are equally sampled without any gaps. The second positional argument \texttt{y} is mandatory and corresponds to the input time series values.
The \texttt{n\_out} argument is a mandatory keyword argument that defines the number of output values\footnote{It is important to note that if there are gaps in the time series (index), fewer than \texttt{n\_out} indices may be returned, as no data points can be selected for empty bins}.
In addition to these arguments, optional keyword arguments can be passed via \texttt{**kwargs}. These additional arguments provide increased flexibility, including options such as the \textit{parallel} argument, a boolean that enables multi-threading when set to \texttt{True}. By default, the \textit{parallel} option is set to \texttt{False}.
Lastly, the \texttt{downsample} method returns a \texttt{numpy} array containing unsigned 64-bit integers, representing the indices of the downsampled (i.e., selected) values.

\section{Illustrative example}\label{sec:example}
Listing~\ref{listing:example} provides an illustration of how \tsdownsample{} can be utilized. 
The code snippet begins by importing the \texttt{MinMaxLTTBDownsampler} class from the \tsdownsample{} package, along with the \texttt{numpy} library. 
\texttt{numpy} is utilized to generate a random time series dataset consisting of 10 million points.
Next, an instance of the \texttt{MinMaxLTTBDownsampler} class is constructed, which is used to downsample the aforementioned time series data to 1000 points. This is accomplished by utilizing the \texttt{downsample} method, whose interface is detailed in Section~\ref{sec:downsampling_interface}. The resulting indices of the downsampled time series data are stored in the \texttt{s\_ds} variable.
Subsequently, these selected indices can be used to retrieve a representative subset of the original time series data, facilitating efficient visualization.

\begin{listing}[ht]
\begin{lstlisting}[language=Python]
from tsdownsample import MinMaxLTTBDownsampler
import numpy as np

# Create a time series
y = np.random.randn(10_000_000)

# Downsample to 1000 points
# ----------------------------------
# Get the selected indices when downsampling y to 1000 points
s_ds = MinMaxLTTBDownsampler().downsample(y, n_out=1000)
\end{lstlisting}
\caption{Downsampling a random array with MinMaxLTTB.}
\label{listing:example}
\end{listing}

\subsection{Integration in plotly-resampler}
\tsdownsample{} has been integrated as downsampling back end in the \texttt{plotly-resampler} visualization tool since version \textit{v0.9}~\cite{vanderdonckt2022plotly_resampler}. At the time of writing, \texttt{plotly-resampler} has over 1.5 million installations.
One longstanding challenge with the \texttt{plotly-resampler} library was the need for users to compile downsampling C code locally during the installation process. This requirement often led to complications, as it necessitates users to have the correct Python headers, ensure compatibility of the \texttt{numpy} version with the utilized C API, and have an appropriate C compiler installed. However, by adopting \tsdownsample{}, these issues have been effectively addressed. \tsdownsample{} has a precompiled binary available for multiple platforms, eliminating the need for users to compile the underlying code themselves, while ensuring optimal performance. This integration has not only resolved these compilation-related obstacles but has also resulted in remarkable speed improvements, ranging from 3 to 30 times faster performance.

% This integration addressed several problems associated with locally compiling downsampling C code during the library installation procedure. In particular, users had proble
% It also resulted in a significant speed improvement, ranging from 3 to 30 times faster. This enhancement was made possible by distributing \tsdownsample{} as a pre-compiled binary for multiple platforms, ensuring optimal performance.

% This integration resolved multiple issues related to locally compiling downsampling C code during the installation process of the library and facilitated a 3-10x+ speedup, since \tsdownsample{} is distributed as a pre-compiled binary for various platforms without sacrificing performance. At the time of writing, \texttt{plotly-resampler} has over 1 million installs.

\section{Performance}\label{sec:benchmarks}
We analyzed the performance of \tsdownsample{} for a range of data types and algorithms, as presented in Table~\ref{tab:benchmarks}. To do so, we created an array of random values for the data types under consideration and measured the time required to downsample the respective array to 2,000 values (i.e., \texttt{n\_out}=2,000) using Python's \texttt{timeit} module. A reproducible notebook containing the benchmark code can be found here: \url{https://github.com/predict-idlab/tsdownsample/blob/main/notebooks/benches.ipynb}.

The benchmarks were executed on a server with an \textit{Intel Xeon E5-2650 v2 (32) @ 3.40GHz} CPU and {SAMSUNG M393B1G73QH0-CMA DDR3 1600MT/s} RAM, running on the \textit{Ubuntu 18.04.6 LTS x86\_64} operating system. Other running processes were limited to a minimum.

%%% TABEL

Table~\ref{tab:benchmarks} displays the median time measurements for the five downsampling algorithms available in \tsdownsample{}. These measurements are provided for all supported data types and varying numbers of data points.
Among the algorithms, EveryNth exhibits a constant execution time of approximately 0.02 ms, regardless of the length of the input data. 
For the M4, MinMax, and MinMaxLTTB algorithms, we observe two main trends. Firstly, there is a linear or sublinear increase (approximately 10x or less) in runtime when transitioning from 10 million to 100 million and then to 1 billion data points. Secondly, these three algorithms have similar runtimes for the same data type, which is most noticeable in the 1 billion data point rows. Both these trends hold true for the sequential and parallel executions. 
In the case of the LTTB algorithm, we again notice a linear scaling pattern as the data length increases. It is important to note that the runtime of LTTB is significantly slower, up to two orders of magnitude, compared to MinMaxLTTB, especially when dealing with larger datasets (e.g., 1 billion uint8 data points). This slower runtime can be attributed to LTTB requiring much more computationally expensive calculations~\cite{steinarsson2013downsampling}, which is largely mitigated in MinMaxLTTB~\cite{van2023minmaxlttb}.

\begin{table}
\resizebox{\columnwidth}{!}{
\begin{tabular}{ll|r|rr|rr|rr|r}
 \toprule
 & \textit{Algorithm} & \multicolumn{1}{c|}{\textbf{EveryNth}} & \multicolumn{2}{c|}{\textbf{M4}} & \multicolumn{2}{c|}{\textbf{MinMax}} & \multicolumn{2}{c|}{\textbf{MinMaxLTTB}} & \multicolumn{1}{c|}{\textbf{LTTB}} \\
 & \textit{Parallel} & \multicolumn{1}{c|}{False} & False & True & False & True & False & True & \multicolumn{1}{c|}{False} \\ \midrule
dtype & N &  &  &  &  &  &  &  &  \\ \bottomrule
\multirow[c]{4}{*}{float16} & 1,000,000 & 0.02 & 0.47 & 0.43 & 0.43 & 0.55 & 1.12 & 0.80 & 6.60 \\
 & 10,000,000 & 0.01 & 2.94 & 0.59 & 2.31 & 0.56 & 4.13 & 0.89 & 59.95 \\
 & 100,000,000 & 0.03 & 24.93 & 4.94 & 34.49 & 4.97 & 25.89 & 5.25 & 575.40 \\
 & 1,000,000,000 & 0.01 & 255.13 & 44.61 & 250.30 & 44.92 & 262.83 & 45.06 & 5614.58 \\
 \midrule
\multirow[c]{4}{*}{float32} & 1,000,000 & 0.01 & 0.41 & 0.27 & 0.46 & 0.23 & 1.17 & 0.56 & 2.41 \\
 & 10,000,000 & 0.03 & 4.02 & 0.89 & 3.56 & 0.94 & 6.29 & 1.15 & 18.39 \\
 & 100,000,000 & 0.02 & 41.27 & 9.25 & 33.40 & 9.28 & 40.40 & 9.56 & 173.72 \\
 & 1,000,000,000 & 0.02 & 407.50 & 88.64 & 338.46 & 88.75 & 398.32 & 89.17 & 1750.17 \\
 \midrule
\multirow[c]{4}{*}{float64} & 1,000,000 & 0.02 & 0.73 & 0.33 & 0.75 & 0.34 & 1.52 & 0.53 & 2.33 \\
 & 10,000,000 & 0.02 & 8.60 & 2.04 & 8.80 & 2.18 & 10.51 & 2.30 & 18.25 \\
 & 100,000,000 & 0.02 & 85.43 & 18.06 & 83.51 & 18.09 & 86.07 & 18.37 & 196.97 \\
 & 1,000,000,000 & 0.01 & 832.17 & 176.74 & 663.86 & 176.82 & 828.15 & 177.26 & 1804.99 \\
 \midrule
\multirow[c]{4}{*}{int8} & 1,000,000 & 0.03 & 0.38 & 0.43 & 0.46 & 0.45 & 0.93 & 0.74 & 3.40 \\
 & 10,000,000 & 0.02 & 1.98 & 0.51 & 1.50 & 0.52 & 2.64 & 0.77 & 24.39 \\
 & 100,000,000 & 0.01 & 18.08 & 2.69 & 14.49 & 2.77 & 17.58 & 3.03 & 237.54 \\
 & 1,000,000,000 & 0.01 & 151.21 & 22.63 & 142.52 & 22.61 & 155.97 & 22.83 & 2374.65 \\
 \midrule
\multirow[c]{4}{*}{int16} & 1,000,000 & 0.02 & 0.38 & 0.35 & 0.41 & 0.37 & 0.86 & 0.64 & 3.34 \\
 & 10,000,000 & 0.01 & 3.78 & 0.46 & 2.01 & 0.50 & 3.41 & 0.74 & 28.64 \\
 & 100,000,000 & 0.02 & 30.17 & 4.95 & 23.65 & 5.07 & 23.87 & 5.20 & 253.98 \\
 & 1,000,000,000 & 0.02 & 230.80 & 44.80 & 229.14 & 44.80 & 232.49 & 45.00 & 2417.14 \\
 \midrule
\multirow[c]{4}{*}{int32} & 1,000,000 & 0.03 & 0.59 & 0.38 & 0.64 & 0.40 & 1.28 & 0.63 & 3.19 \\
 & 10,000,000 & 0.02 & 5.36 & 0.92 & 4.25 & 1.01 & 5.05 & 1.21 & 23.79 \\
 & 100,000,000 & 0.01 & 44.57 & 9.61 & 45.47 & 9.31 & 58.25 & 9.78 & 227.60 \\
 & 1,000,000,000 & 0.01 & 452.85 & 88.67 & 390.95 & 88.72 & 470.03 & 89.00 & 2297.82 \\
 \midrule
\multirow[c]{4}{*}{int64} & 1,000,000 & 0.03 & 1.52 & 0.45 & 1.55 & 0.57 & 1.86 & 0.70 & 3.25 \\
 & 10,000,000 & 0.01 & 13.33 & 2.20 & 11.15 & 2.28 & 11.55 & 2.50 & 24.63 \\
 & 100,000,000 & 0.03 & 116.66 & 18.50 & 139.79 & 18.17 & 134.90 & 19.30 & 270.74 \\
 & 1,000,000,000 & 0.03 & 1154.85 & 177.20 & 1114.58 & 177.47 & 1171.36 & 177.52 & 2408.60 \\
 \midrule
\multirow[c]{4}{*}{uint8} & 1,000,000 & 0.02 & 0.39 & 0.44 & 0.47 & 0.48 & 0.91 & 0.74 & 3.42 \\
 & 10,000,000 & 0.02 & 2.08 & 0.52 & 1.61 & 0.53 & 2.65 & 0.77 & 24.52 \\
 & 100,000,000 & 0.02 & 15.97 & 2.70 & 15.01 & 2.78 & 18.86 & 2.96 & 243.15 \\
 & 1,000,000,000 & 0.01 & 155.68 & 22.73 & 147.67 & 22.62 & 160.77 & 22.79 & 2384.09 \\
 \midrule
\multirow[c]{4}{*}{uint16} & 1,000,000 & 0.02 & 0.52 & 0.34 & 0.37 & 0.35 & 0.81 & 0.55 & 3.29 \\
 & 10,000,000 & 0.01 & 2.56 & 0.45 & 2.02 & 0.45 & 3.22 & 0.62 & 24.86 \\
 & 100,000,000 & 0.01 & 24.31 & 4.88 & 20.95 & 4.82 & 24.79 & 4.89 & 317.30 \\
 & 1,000,000,000 & 0.01 & 239.21 & 44.52 & 202.76 & 44.70 & 238.73 & 44.74 & 2392.48 \\
 \midrule
\multirow[c]{4}{*}{uint32} & 1,000,000 & 0.02 & 0.58 & 0.34 & 0.62 & 0.36 & 1.15 & 0.57 & 3.30 \\
 & 10,000,000 & 0.02 & 5.98 & 0.92 & 4.33 & 0.93 & 8.88 & 1.16 & 25.02 \\
 & 100,000,000 & 0.01 & 47.35 & 9.31 & 40.29 & 9.27 & 48.17 & 9.56 & 240.22 \\
 & 1,000,000,000 & 0.01 & 468.04 & 88.72 & 399.56 & 88.91 & 475.70 & 88.91 & 2402.57 \\
 \midrule
\multirow[c]{4}{*}{uint64} & 1,000,000 & 0.01 & 1.85 & 0.40 & 1.87 & 0.42 & 2.14 & 0.61 & 3.77 \\
 & 10,000,000 & 0.01 & 14.69 & 2.11 & 13.90 & 2.15 & 14.17 & 2.42 & 35.38 \\
 & 100,000,000 & 0.02 & 140.70 & 18.19 & 137.31 & 18.21 & 140.87 & 18.36 & 276.40 \\
 & 1,000,000,000 & 0.01 & 1403.84 & 176.99 & 1380.90 & 177.19 & 1402.95 & 177.25 & 2817.33 \\
 \bottomrule
\end{tabular}
}
\label{tab:benchmarks}
\caption{Downsampling time (in ms) for the algorithms offered by \tsdownsample{}. The experiment parameters are described in the first two columns: \textit{dtype} indicates the data type, and \textit{N} denotes the number of data points. Note that LLTB cannot be parallelized, since this algorithm requires a sequential iteration over the bins~\cite{van2023minmaxlttb}.}
\end{table}

% Findings
% EveryNth has constant time, independend of the data length - it takes around 0.02 ms.
% For M4, MinMax, and MinMaxLTTB we observe as first trend a linear (i.e., a ca. 10x increase) or even sublinear (i.e., a smaller than 10x increase) when going from 10M to 100M and to 1B data points. 
% As second trend, the runtime for these three algorithms is generally +/- the same for most data types. 
% Both trends are true for the sequantial and parallel execution of these three algorithms.
% Finally, for the LTTB algorithm we observe again a linear scaling with increasing data length. Note however, that the runtime of LTTB is up to 2 orders of magnitude slower than that of MinMaxLTTB (e.g., 1B uint8 datapoints).

%%% FIGUUR

\begin{figure}[htpb]
    \centering
    \includegraphics[width=\textwidth]{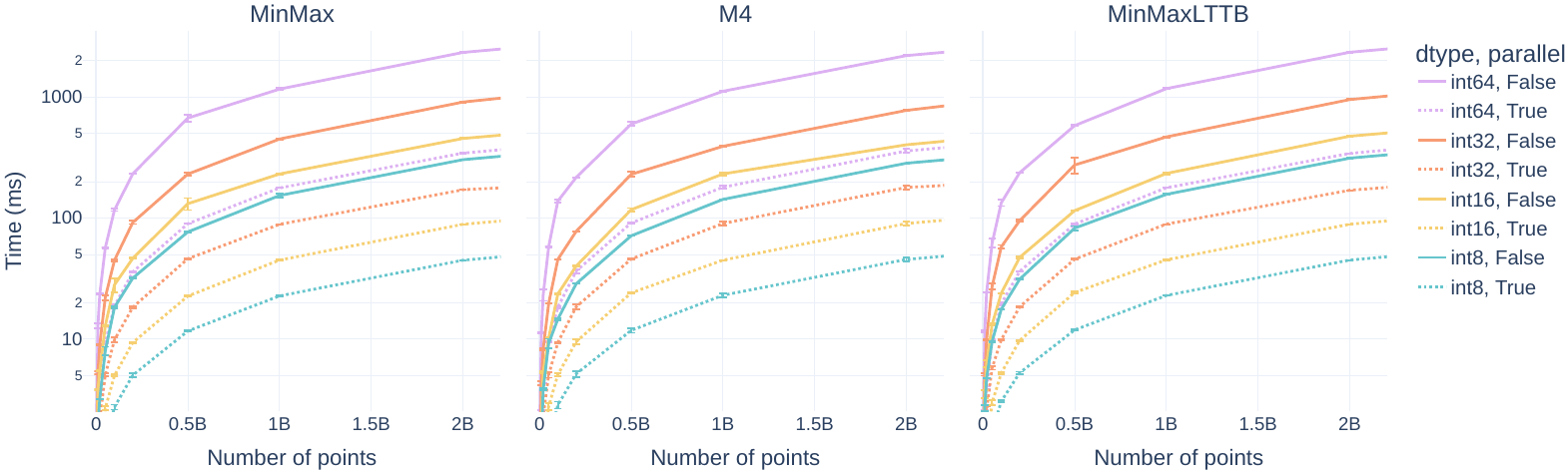}
    \caption{Downsampling time comparison for integer data types for the MinMax, M4, and MinMaxLTTB algorithms (shown in the three subplots) provided by \tsdownsample{}. The y-axis represents the downsampling time in milliseconds on a logarithmic scale, while the x-axis indicates the number of data points. 
    The figure includes (nearly imperceptible) whiskers to denote the standard deviation of the measurements for the collected data points.
    }
    \label{fig:benchmarks}
\end{figure}

Figure~\ref{fig:benchmarks} provides additional insights to complement the findings of Table~\ref{tab:benchmarks}. It illustrates the relationship between downsampling time (y-axis) and the number of data points (x-axis) for the three algorithms that utilize the optimizations of ArgMinMax. Given that the y-axis is logarithmic scale, the logarithmic trend that we observe for all integer data types in every subplot, confirms the earlier mentioned observation that the implementation scales linearly with the number of data points in the array. 

Our primary finding is that the implementation exhibits faster performance for lower bitsize variants of the same data type. For instance, int32 demonstrates a roughly 2x speed improvement compared to int64 for the same number of data points, while int16 is 2x faster than int32. Remark that these 2x performance differences are even more pronounced (i.e., a clear 2x) for the parallel execution.
This discrepancy in performance can be attributed to the fact that reducing the bit-representation by 2x (e.g., int32 vs. int64) allows for a 2x increase in the number of values that fit in the CPU's SIMD registers. This utilization of SIMD registers is an essential part of the ArgMinMax code base and results in fewer read (\texttt{memcpy}) instructions, which impacts performance since ArgMinMax is primarily bound by memory access. %Thus, the number of (read) instruction calls is roughly halved.
% It is worth noting that the speedup achieved for int8 vs. int16 is only 1.4x, rather than 2x, owing to the considerably higher cost associated with (much more frequent) additional bookkeeping required for overflow-free code. This includes tasks such as exiting the core loop and performing (much slower) horizontal SIMD operations.

Our second key finding emphasizes the benefits of implementing multithreading. On average, multithreading leads to an impressive 7x performance improvement on the benchmarking computer. Notably, when extrapolating the linear trend of the int64 data type, \tsdownsample{} demonstrates the capability to downsample data at a rate of 45 GB/s (i.e., 8 GB / 0.177 s).

\section{Conclusion}
Time series visualization plays a crucial role in exploratory data analysis, particularly as datasets continue to grow in size. To enable scalable line-chart visualization, downsampling has emerged as a well-established technique. However, we have observed a need for a convenient and high-performance time series downsampling solution within the Python landscape, facilitating the integration of downsampling capabilities into widely used Python visualization packages.
To address this need, we introduce \tsdownsample{}, a Python library that offers a convenient interface to leading precompiled downsampling algorithms, harnessing highly optimized underlying Rust code. A key aspect of achieving high performance in \tsdownsample{} involved optimizing argmin and argmax operations using SIMD instructions and leveraging multithreading. The runtime feature set detection enables the selection of the most optimal implementation based on the CPU, allowing for the distribution of a single binary that can cater to multiple CPU feature sets within the same architecture.
Benchmark results confirm the critical role played by both SIMD optimizations and multithreading in achieving the impressive performance of \tsdownsample{}. We firmly believe that \tsdownsample{} represents a significant advancement in delivering high-performance time series downsampling capabilities to the Python ecosystem. This advancement is further illustrated by the adoption of \tsdownsample{} in the \texttt{plotly-resampler} tool, solidifying its position within the Python community.

\section*{Acknowledgements}
The authors thank Martijn Courteaux and Tom Windels for having fruitful discussions on writing efficient low-level code.

%% References:
\bibliographystyle{elsarticle-num} 
\bibliography{references}

\section*{Current code version}
\label{}

% Ancillary data table required for subversion of the codebase. Kindly replace examples in right column with the correct information about your current code, and leave the left column as it is.

\begin{table}[!h]
\begin{tabular}{|l|p{6.5cm}|p{6.5cm}|}
\hline
\textbf{Nr.} & \textbf{Code metadata description} & \textbf{Please fill in this column} \\
\hline
C1 & Current code version & v0.1.2 \\
\hline
C2 & Permanent link to code/repository used for this code version & \url{https://github.com/predict-idlab/tsdownsample/releases/tag/v0.1.2} \\
\hline
C3  & Permanent link to Reproducible Capsule & Not available \\
\hline
C4 & Legal Code License   & MIT\\
\hline
C5 & Code versioning system used & git \\
\hline
C6 & Software code languages, tools, and services used & Python, Rust \\
\hline
C7 & Compilation requirements, operating environments \& dependencies & \texttt{cargo} is used to manage the Rust dependencies and compilation. \href{https://github.com/PyO3/maturin}{PyO3/maturin} is used as build backend to generate Python bindings for the Rust code. \\
\hline
C8 & If available Link to developer documentation/manual & Not available \\
\hline
C9 & Support email for questions & jeroen.vanderdonckt@ugent.be \\
\hline
\end{tabular}
\caption{Code metadata (mandatory)}
\label{} 
\end{table}

\section*{Current executable software version}
\label{}

% Ancillary data table required for sub version of the executable software: (x.1, x.2 etc.) kindly replace examples in right column with the correct information about your executables, and leave the left column as it is.

\begin{table}[!h]
\begin{tabular}{|l|p{6.5cm}|p{6.5cm}|}
\hline
\textbf{Nr.} & \textbf{(Executable) software metadata description} & \textbf{Please fill in this column} \\
\hline
S1 & Current software version & v0.1.2 \\
\hline
S2 & Permanent link to executables of this version  & \url{https://github.com/predict-idlab/tsdownsample} \\
\hline
S3  & Permanent link to Reproducible Capsule & Not available \\
\hline
S4 & Legal Software License & MIT \\
\hline
S5 & Computing platforms/Operating Systems & For example Android, BSD, iOS, Linux, OS X, Microsoft Windows, Unix-like , IBM z/OS, distributed/web based etc. \\
\hline
S6 & Installation requirements \& dependencies & Python 3.7+, Rust nightly \\
\hline
S7 & If available, link to user manual - if formally published include a reference to the publication in the reference list & Not available \\
\hline
S8 & Support email for questions & jeroen.vanderdonckt@ugent.be \\
\hline
\end{tabular}
\caption{Software metadata (optional)}
\label{} 
\end{table}

\end{document}